\documentclass[review]{elsarticle}

\usepackage{lineno,hyperref}
\modulolinenumbers[5]

\journal{International Journal of Disaster Risk Reduction}

\usepackage{epstopdf}
\usepackage{subfigure}
\usepackage{multirow}
\usepackage{multicol}
\usepackage{makecell}
\usepackage[table]{xcolor}
\usepackage{fancyhdr}
\usepackage{hyperref}
\usepackage{mathtools}
\usepackage{booktabs}

\begin{document}



\begin{frontmatter}

\title{A multi-modal approach towards mining social media data during natural disasters - a case study of Hurricane Irma}



\author[a1]{Dr. Somya D. Mohanty\corref{mycorrespondingauthor}}
\cortext[mycorrespondingauthor]{Corresponding author}
\ead{sdmohant@uncg.edu}
\author[a1]{Brown Biggers}
\author[a1]{Saed Sayedahmed}
\author[a2]{Nastaran Pourebrahim}
\author[a2]{Dr. Evan B. Goldstein}
\author[a2]{Dr. Rick Bunch}
\author[a3]{Dr. Guangqing Chi}
\author[a1]{Dr. Fereidoon Sadri}
\author[a4]{Dr. Tom P. McCoy}
\author[a5]{Dr. Arthur Cosby}

\address[a1]{Department of Computer Science, University of North Carolina - Greensboro}
\address[a2]{Department of Geography, Environment, and Sustainability, University of North Carolina - Greensboro}
\address[a3]{Department of Agricultural Economics, Sociology, and Education, Population Research Institute, and Social Science Research Institute, The Pennsylvania State University}
\address[a4]{Department of Family \& Community Nursing, University of North Carolina - Greensboro}
\address[a5]{Social Science Research Center, Mississippi State University}
\address[b1]{167 Petty Building, Greensboro, NC 27402-6170}




\begin{abstract}
Streaming social media provides a real-time glimpse of extreme weather impacts. However, the volume of streaming data makes mining information a challenge for emergency managers, policy makers, and disciplinary scientists. Here we explore the effectiveness of data learned approaches to mine and filter information from streaming social media data from Hurricane Irma’s landfall in Florida, USA. We use 54,383 Twitter messages (out of 784K geolocated messages) from 16,598 users from Sept. 10 - 12, 2017 to develop 4 independent models to filter data for relevance: 1) a geospatial model based on forcing conditions at the place and time of each tweet, 2) an image classification model for tweets that include images, 3) a user model to predict the reliability of the tweeter, and 4) a text model to determine if the text is related to Hurricane Irma. All four models are independently tested, and can be combined to quickly filter and visualize tweets based on user-defined thresholds for each submodel. We envision that this type of filtering and visualization routine can be useful as a base model for data capture from noisy sources such as Twitter. The data can then be subsequently used by policy makers, environmental managers, emergency managers, and domain scientists interested in finding tweets with specific attributes to use during different stages of the disaster (e.g., preparedness, response,  and recovery), or for detailed research.
\end{abstract}


\end{frontmatter}


\section{Introduction}

Climate change is expected to drive increases in the intensity of tropical cyclones \citep{knutson_tropical_2019} and increase the occurrence of ‘blue sky’ flooding \citep{moftakhari_increased_2015}. Despite these hazards, coastal populations \citep{neumann_future_2015}, and investments in the coastal built environment \citep{lazarus_building_2018} are likely to grow. Understanding the impact of extreme storms and climate change on coastal communities requires pervasive environmental sensing. Beyond the collection of environmental data streams such as river gages, wave buoys, and tidal stations, internet connected devices such as mobile phones allow for the creation of real-time crowd-sourced information during extreme events. A key area of research is understanding how to use streaming social media information during extreme events --- to detect disasters, provide situation awareness, understand the range of impacts, and guide disaster relief and rescue efforts \citep[e.g.][]{de_longueville_omg_2009, sakaki_earthquake_2010,tyshchuk_social_2012,middleton_real-time_2014,muralidharan_hope_2011,imran2020rapid, niles2019social}. 

Twitter -- with approximately 600 million tweet posts every day \citep{livestats_internet_2014} and programmatic access to messages -- has become one of the most popular social media platforms, and a common data source for research on extreme events \citep[e.g.][]{de_bruijn_global_2019,kryvasheyeu2016rapid, pourebrahim_understanding_2019}.  In addition to text,  a subset of messages shared across Twitter contain images captured by its users (20 - 25\% of messages contain images / videos \citep{leetaru_visualizing_2019}). A key hurdle for studying these aspects of extreme events with Twitter is the data are both large and considerably noisier than curated sources such as dedicated streams of information (e.g., dedicated environmental sensors). Posts on Twitter during disasters might also be irrelevant, or provide mis- or dis- information \citep[e.g.,][]{gupta2013faking,laylavi_event_2017}, highlighting the importance of filtering and subsetting social media data when used during disaster events. Therefore a key step in all work with Twitter data is to filter and subset the data stream.

Previous work has addressed filtering and subsetting Twitter data during hazards and other extreme events. Techniques have included relying on specific hashtags \citep[e.g.,][]{murzintcev2017disaster},  semantic filtering \citep{abel_semantics_2012}, keyword-based filtering \citep{laylavi_event_2017}, as well as natural language processing (NLP) and text classification that use machine learning algorithms \citep{laylavi_event_2017,liu_assessing_2019}.  Classifiers such as support vector machine and Naive Bayes classifiers have been used to differentiate between real-world event messages and non-event messages \cite{becker_beyond_2011}, and to extract valuable ``information nuggets'' from Twitter text messages\cite{imran_extracting_2013}. The tweets’ length and textual features can also used to filter emergency-related tweets \cite{imran_extracting_2013}. Tweets have been scored against classes of event-specific words (term-classes) to aid in filtering \cite{laylavi_event_2017}.  Previous work have filtered and subset tweets using expert-defined features of the tweet \citep{zahra2020automatic}. Images have also been used to subset tweets based on the presence/absence of visible damage \citep{ilyas2014microfilters}. Filtering can also be understood by the extensive work on determining the relevance of tweets for a given event — see recent work and reviews \citep{kaufhold2020rapid, sit2019identifying,imran2020using}. In the context of this paper, we view filtering as any generic process that subsets tweets, even beyond the binary class division of relevance.

A few studies have identified the significance of adding spatial features and external sources for a better assessment of tweets’ relevance for disaster events. For example, \cite{spinsanti_automated_2013} enriched their model with geographic data to identify relevant information. Previous work has used spatio-temporal data to determine tweet relevance \cite{liu_assessing_2019}, or linked geolocated tweets to other environmental data streams \citep[e.g.,][]{de2015geographic, de2020improving}.

As observed from prior work, capturing situational awareness information from social media data involves a hierarchical filtering approach \citep{landwehr2014social}. Specifically, researchers/interested stakeholders filter down the data from the noisy social media data stream to fit their specific use cases (such as, type of image - destruction, damage, flooding; type of text - damage, donation, resource request/offer; spread of information, etc). A key component in such an approach is the quality of baseline data capture. Towards this our study proposes a novel approach towards quality gating the data capture from the social media data streams using developed threshold measures. This baseline filtering methodology that can be used to find relevant tweets and refining the data capture routines. Specifically, the goal of our study is to explore a multi-modal filtering approach which can be used to provide situational awareness from social media data during disaster events. We develop an initial prototype using tweets from Florida, USA during Hurricane Irma. The filtering routine allows users to adjust four separate models to filter Twitter messages: a geospatial model, an image model, a user model, and a text model. All four models are tested separately, and can be operated independently or in tandem. This is a design feature as we envision the sorting and filtering thresholds will be different for different users, for different events, and for different locations. We work through each model and discuss the combined model in the following sections.

\section{Methodology}

\subsection{Hurricane Irma}

\begin{figure}[ht!]
\centering
\includegraphics[width=\textwidth]{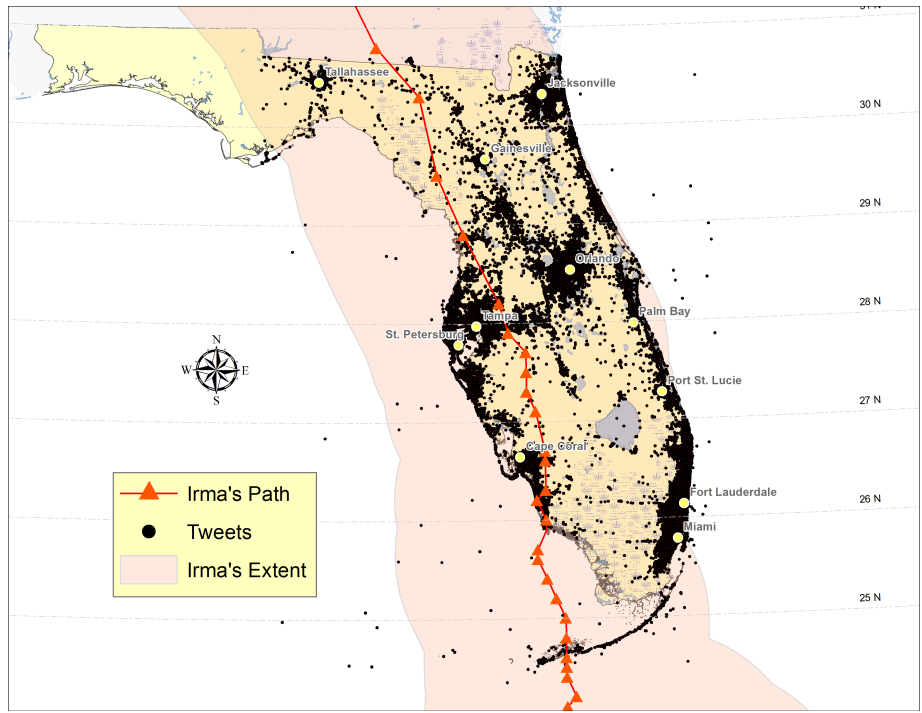}
\caption{The path of Hurricane Irma in September 2017 (orange line), the extent of tropical storm force winds (pink outline), and the location for all 784K geolocated tweets used as the basis for this study (black dots).} 
\label{fig:irma}
\end{figure}

Hurricane Irma (Figure~\ref{fig:irma}) was the first category 5 hurricane in the 2017 Atlantic hurricane season \citep{cangialosi_hurricane_2018}. Hurricane Irma formed on August 31st, 2017, impacting many islands of the Caribbean, and finally dissipating over the continental United States \citep{cangialosi_hurricane_2018}. Here we focus on the Twitter record of Irma specifically in Florida, USA. Irma made landfall in Florida Keys on 09/10/2017 as a category 4 hurricane and dissipated shortly after 09/13/2017. 134 fatalities were recorded as a result of the hurricane, with an estimated loss of \$64.76 billion \citep{nguyen_forecasting_2019}, making it one of the costliest hurricanes in the history of the United States \citep{united_states_national_hurricane_center_costliest_2018}. 

\subsection{Data collection and preprocessing}

\subsubsection{Twitter data}
We used the Twitter Application Programming Interface (API) to collect tweets located in the geospatial bounding box that captured the state boundary of Florida. Tweets were recorded for the period of 09/01/2017 to 10/10/2017, and resulted in the collection of 784K tweets from 96K users during the time period. Our work is focused on 72 hours (09/10/2017 - 09/12/2017) when Hurricane Irma was near or over Florida. Therefore we subset the data and use 54,383 tweets from 16,598 users during this 72hr window. Figure 1 highlights the locations of the Twitter messages, along with the path of Hurricane Irma, and the extent of tropical Storm force winds.

Each tweet from the Twitter API has 31 distinct metadata attributes \citep{twitter_introduction_2019} that can conceptually be grouped into three categories: 1) Spatio-temporal (time of creation and geolocation [latitude, longitude]), 2) Tweet content (tweet text, weblinks, hashtags, and images), and 3) Tweet source (account age, friends count, followers count, statuses count, and if verified). Geolocated tweets can have one of two types of location data — Places or Coordinates. Coordinates are exact locations with latitude and longitude attributes, while Places are locations within a Bounding Box or a Polygon designating a geospatial area in which the tweet is recorded \citep{twitter_geo_2019}. For tweets with Places attributes, we transform the area representation to a single point by selecting the centroid of the Polygon as the location represented by the tweet. Within our study 42.58\% (23,157) of the tweets had Coordinate locations and 57.42\% (31,226) had Place locations.

\subsubsection{Geospatial data}
\label{sec:geomodel}

We collected meteorological sensor data, wind speed (in mph) and precipitation (in inches), for each county in Florida for the 72 hours (09/10/2017 - 09/12/2017). The hourly wind speeds was collected from the NOAA National Centers for Environmental Information (NCEI). Hourly precipitation values were obtained from the United States Geological Survey’s Geo Data Portal (USGS GDP) of the United States Stage IV Quantitative Precipitation Archive. Precipitation values from the closest weather station were used due to difficulty in obtaining reliable data for all weather stations.  In addition to meteorological forcing, we collected data consisted of location of the hurricane’s eye, category of the hurricane, pressure and wind speed (NOAA National Hurricane Center). This data were discretized into hourly windows for the 72 hours. 

\subsubsection{Data pre-processing}
We aligned the 72hrs of Twitter data and the corresponding 72hrs of meteorological forcing data.  Wind and precipitation values at the geolocation of a tweet was calculated using Inverse Distance Weighting (IDW). IDW is an interpolation method that calculates a value at a certain point using values of other known points:

\begin{equation}
    W_p = \frac{\sum_{i=1}^{n}{\frac{W_i}{D_{i}^{k}}}}{\sum_{i=1}^{n}{\frac{1}{D_{i}^{k}}}} 
\end{equation}{}

where, $W_p$ is the wind speed to be interpolated at point $p$, $W_i$ is the wind speed at point $i$, $D_i$ is the distance between point $p$ and $i$, and $k$ is a power function that reduces the effect of distant points as it goes up. IDW has been widely used to interpolate climatic data \citep{tomczak1998spatial}. The method has demonstrated accurate results when compared to other interpolation methods especially in regions characterized by strong local variations \citep{yang2015spatial}. IDW, for example, assumes that any measurement taken at a fixed location (e.g., weather station) has local influence on surrounding area, and the influence decreases with increasing distance. Within our study we chose IDW as our interpolation method as meteorological factors in a hurricane are highly influenced by local variations.

Furthermore, each tweet was also annotated with the corresponding temporal hurricane conditions data. Specifically, for each hourly time window, a tweet was associated with its distance from the eye of the hurricane and its conditions (i.e., pressure, max wind speed) during that window.

\subsection{Multimodal scoring of tweet relevance}

Our goal is to develop a single model for tweet relevance based on four sub models — 1) the relevance of the tweet based on \textit{geospatial} attributes (i.e., the Tweets location relative to the forcing conditions of the hurricane, 2) the relevance of tweet \textit{images} (when media is included in the tweet), 3) a score for the reliability of the \textit{user} (i.e., network attributes to predict if a user is ‘verified’ by Twitter),  and 4) the relevance of the tweet \textit{text}. The methods used to construct of each of these models, and the results of models (submodels and the combined model) in are discussed in Section 3.

\subsubsection{Geospatial model}

Our goal in designing a geospatial relevance model was to search for thresholds in forcing conditions where tweets were likely to be related to Hurricane Irma, as opposed to background social media discussions. Specifically, we posit that the messages which are in close geospatial and temporal proximity to the disaster event will have more relevant situational awareness information than those which are not. Furthermore, such an approach can be used in real-time during the occurrence of an event where meteorological data can provide key information about disaster's impact at different locations. 

There are many meteorological conditions that can be used as proxies for extreme disaster conditions. We focus here on searching for modeling functions relating wind speed ($w$), precipitation ($p$), and distance from hurricane eye ($d$). We acknowledge that other factors could be used in addition to these three attributes. For example, rainfall during a given interval could be quantified in several ways, such as mean rainfall rate, max rainfall rate, total rainfall in a given interval. Similarly Wind metrics could include mean wind speed, max wind speed, metrics based on wind gusts, etc.. For locations nearby the coast, metrics could include tide elevations, or storm surge elevations, and locations near streams could include stage and discharge data. Ultimately we chose Wind speed, precipitation and distance from the hurricane eye as these factors are available everywhere (vs metrics that are only applicable along streams and rivers) and because they are commonly available and collected by even basic meteorological stations. Nine different functions, --- 1) $\frac{wind\ *\ rain}{distance}$, 2) $\frac{rain}{distance}$, 3) $\frac{wind}{distance}$, 4) $\frac{wind\ *\ rain}{\sqrt[]{distance}}$, 5) $\frac{rain}{\sqrt[]{distance}}$, 6) $\frac{wind}{\sqrt[]{distance}}$, 7) $\frac{wind\ *\ rain}{\sqrt[3]{distance}}$, 8) $\frac{rain}{\sqrt[3]{distance}}$, and 9) $\frac{wind}{\sqrt[3]{distance}}$, combining the geospatial attributes were compared  to identify the best suited model towards creating a relevance score for the tweets. In each of the models, wind speed and precipitation acted as numerators (individually or combined), where as distance was used as a denominator — this was a heuristic method, as tweets are likely more relevant if forcing conditions are more severe (Higher wind, more precipitation, closer distance to hurricane)

%

Approximately 19,000 tweets from the Irma dataset were hand labeled by human coders as “Irma related” or “non-Irma related” based on the tweet content. The performance of each geospatial function was evaluated by comparing the ratio of Irma related tweets to total number of tweets during each time window. The ratios obtained from each formula was normalised using three different approaches - Min-max scaling, Log ($log_{10}()$), and Box-Cox transformations. Ranking of Shapiro-Wilk (SW) test statistics was used to assess normality. In addition, multiple observed statistics of mean, standard deviations, and percentage of values within 1, 2, and 3 standard deviations from the mean were calculated to evaluate normality.  The goal of this normalization procedure was to establish a comparative scoring range for each of the models. The scores enable development of a combined overall model for filtering tweets relevant to the hurricane (as described in Section~\ref{sec:overall}. Apart from the ratio, we also evaluated the F1-score ($F1 = 2\frac{precision*recall}{precision + recall}$) for the model, where $precision = \frac{TP}{TP + FP}$, $recall = \frac{TP}{TP + FN}$, and $TP$, $FP$, $TN$, and $FN$ represent the number of true positives, false positives, true negatives, and false negatives, respectively.

\subsubsection{Image model}
Supervised machine learning models were used to develop automated image classification of images in the Twitter dataset. The goal of this model is two fold:  1) to develop, a binary classifier capable of distinguishing hurricane-related images from the non-related ones, and 2) to then develop a multi-label annotator capable of classifying the hurricane-related images into one or more of three incident categories --- 1) Flood, 2) Wind, and 3) Destruction. 

A key hurdle in the approach was the lack of available labeled training data for supervised classification. We developed a web platform for image labeling for annotation by human coders. The platform took unlabeled images, and displayed them on a browser for human coders to annotate. Within the browser, the coder was asked the question of --- \textit{Does this image have any of the following --- 1) Flooding, 2) Windy, and 3) Destruction?} An image is considered ``Flooding'' if there is water accumulation in an area of the image. An image is considered ``Windy'' if there are visual elements in the picture which show tree branches are moving in a direction or some objects that are flying or heavy rain visible in the image. An image is considered to have ``Destruction'' if there is damage to property, vehicles, roads, or permanent structures. An image can be in one or more of the previous classes. If an image has one of the codified classes, it is labelled as Irma \textit{related} and if it does not have any of them the image is labelled as \textit{not related}. 

For the dataset, approximately 7,000 images were labeled by 3 human coders/raters where the data was divided equally between them. Following codification resulting dataset had the following distribution --- Related: 817 / Not-Related: 6081 images, and Wind: 120 images; Flooding: 266 images; Destruction: 571 images. We also evaluated the inter-rater reliability using Light's Kappa \citep{light1971measures} for 100 sampled images (with balanced distribution of related and not-related classes) that were labeled by all three coders. Agreement between all three coders for related versus not-related was at $0.77$, and across tags Flooding - $0.88$; Windy $0.27$; and Destruction $0.78$. This shows significant agreement among the coders on the labeling \citep{mchugh2012interrater} other than the Windy tag (poor/chance agreement).


This annotated dataset was used to train deep learning models based on convolutional neural network (CNN) architectures. Convolutional networks have been widely used in large-scale image and video recognition \citep{simonyan_very_2014}. CNN architecture consists of an input layer, an output layer, and several hidden layers in between.  A hidden layer can be a convolutional layer, a pooling layer (e.g. max, min, or average), a fully connected layer, a normalization layer, or a concatenation layer. Within our approach, we evaluated three modern CNN architectures --- 1)VGGNet, 2)ResNet, and 3)Inception-v3, and compared the performance of each model to its counterparts.

In VGGNet \citep{simonyan_very_2014}, the image is passed through a stack of CNN layers, where filters with a very small receptive field is used. Spatial pooling is done by five max-pooling layers, which is followed by convolution layers. The limitation of VGGNet is its large number of parameters, which makes it challenging to handle. Residual Neural Network (ResNet) \citep{he_deep_2015} was developed with fewer filters and in turn has lower complexity than VGGNet. While the baseline architecture of ResNet was mainly inspired by VGGNet, a shortcut connection was added to each pair of filters in the model. In comparison, Inception-v3 \citep{szegedy_rethinking_2015} uses convolutional and pooling layers which are concatenated to create a time/memory efficient architecture. After the last concatenation, a dropout layer is used to drop some features to prevent overfitting before proceeding with final result. The architecture is quite versatile, where it can be used with both low-resolution images and high-resolution images, and can distinguish any size of a pattern, from a small detail to the whole image. This makes it useful in our application as the quality and type of image can vary widely due to disparate smart devices used by the Twitter population. The pre-trained Inception-V3 is trained on the Imagenet \citep{krizhevsky_imagenet_2012} dataset which consists of hundreds of thousands of images in over one thousand categories. The weights of this model are used as a starting point for training and fine tuned using our sample images. The approach takes advantage of transfer learning \citep{oquab_learning_2014}, where the classifier is able to initially learn features of physical objects in a wide variety of scenarios and then trained on specific observations within our data. This enables a more accurate and generalizable model.

Data augmentation methods were used to expand the number of training samples and therefore improve model accuracy. For example, additional training images are generated by rotating and scaling of the original images. This was done to balance the number of images of Irma related to the un-related ones. The resulting dataset consisted of approximately 6,000 images in each class for the binary classifier, and approximately 2,000 images in each class for the annotator model. The models were trained and testing using a 70-30 split on the dataset. For each model, performance scores (precision, recall, and F1) was recorded. Probability scores for each tweet image were then recorded for every class, which was further normalized using log-transform and re-scaled using min-max scaling to be used in the overall model.

\subsubsection{User model}

It is essential to quantify the authenticity of user accounts which have posted messages and images during a disaster event. For the purpose, our goal was to develop a scoring model which can provide continuous probabilistic measures of account authenticity. 

Manually annotating user reliability in a large dataset such as Twitter is not practical. As we did not have a labeled dataset, our starting point was to consider the user ``Verified'' attribute within the tweets. The ``Verified'' attribute is annotated to user accounts which Twitter defines to be of public interest \citep{twitterverified}. Within our dataset we had 94,445 non-verified users and 1,692 verified users. Since Twitter’s methodology for finding verified accounts is not public, we aim to develop a proxy automated model. The aim here is to create a model which can help identify users who are also likely to be accounts of public interest and authentic, but remain unverified. This can be used in conjunction with the Twitter ``Verified'' accounts to provide a comprehensive source of authentic accounts during a disaster event. Specifically, our approach provides the adjust the authenticity thresholds based on the continuous probabilistic scores of the model, which enables collection information from accounts which have not yet been verified by Twitter but have similar properties to that of a ``Verified'' account.

The automated model was developed based on supervised machine learning. Specifically, machine learning models \citep{kotsiantis_supervised_2007} were developed for binary classification machine  to predict the label user ``Verified'' (true/false) based on the features of tweets content (weblinks, hashtags) and its creator (account age, friends count, followers count, statuses count). Random Forest (RF) \citep{liaw_classication_2002}, Gradient Boosted (GB) \citep{ye_stochastic_2009}, and Logistic Regression (LR) \citep{kleinbaum_logistic_2010} classifiers were used to train and test the model. 




RF is an ensemble model which consists of multiple decision trees trained on the data and their voting to determine the label class of an observation based on the features. A decision tree has a set of rules, when evaluated on an input, it returns a prediction of a class or a value. RF also returns the ratios of votes for each class it is trained on. A Gradient Boosted (GB) is also an ensemble model which builds decision trees leveraging gradient descent to minimize information loss. Similar to RF, GB also uses weighted majority vote of all of the decision trees for classification. In comparison, Logistic Regression (LR) is a non-parametric model which tries to find the best linear model to describe the relationship between independent variables and a binary outcome for classification.

The output of each of the trained binary models is a classifier capable of predicting if a user can be verified or not. The performance of the resulting model was evaluated using a 10-fold cross validation \citep{domingos_few_2012}, with a 70-30 train test split used in each fold. Furthermore, grid search \citep{bergstra_random_2012} was used on the best performing model for hyper-parameter optimization. Grid search takes in a set of values for each hyperparameter (e.g. number of trees in a forest, max depth of a tree, sample splits, max number of leaf nodes, etc.), folds number, and conducts a search using each possible combination of hyperparameters by evaluating them on a scoring metric such as F1-score. The final output of this model is a min-maxed log-transformed value of the probability scores. This was done to reduce the skewness in score distribution needed for the overall model (described later).

\subsubsection{Text model}
\label{sec:textmodel}

The goal of the text model was to delineate tweets with Irma related text from those addressing other topics. While generic search term such as the ``Hurricane Irma'' can provide a starting point, prior research \citep{yarowsky_unsupervised_1995,marco_clustering_2013,arora_linear_2018} in the domain has shown that content organically develops to other words. An automated system trained on a large corpus to recognize context may improve the results, but this suffers from two significant pitfalls. First, training a learner on large bodies of text is costly from the perspective of computational overhead \citep{imran_twitter_2016}. Second, the dynamic nature of discussions during a disaster, especially in a format as compact as Twitter, can alter the most likely interpretation of a word’s meaning, resulting in false positives in the captured tweets \citep{de_boom_semantics-driven_2015}.

In order to address the issues we developed a dynamic word embedding model which utilizes online learning to update its learned context. Specifically, we use a neural network based word embedding architecture - Word2Vec \citep{mikolov_efficient_2013,goldberg_word2vec_2014}, which captures the semantic and syntactic relationships between the words present in tweets corpora. In the Word2Vec module, each word is evaluated based upon its placement among other words within a tweet. This target word, combined with its neighboring words before and after its occurrence in a given tweet, is then given to a neural network whose hidden layer  weights correspond with the vector representing the target word. Once the vectors for each word are generated, the vectors can be compared based upon their cosine similarity. As two words get closer in similarity, the vectors representing those words will become closer within vector space; the angle internal to the vector will get smaller; and the cosine of this angle will get closer to, but not exceed, 1. As a result, the similarity in context between a word and its neighbors in vector space can be compared numerically by looking at the cosine of the internal angle formed by two word vectors \citep{ozdikis_semantic_2012}.

Within our approach, tweets were parsed and grouped into 24-hour segments, with primary testing done on the time period  immediately before and after  the initial landfall. Prior to training the model, tweets were first cleaned to eliminate punctuation, numbers, and extraneous/stop words. Each tweet temporally isolated and parsed into token words, to create input vectors for training and testing of Word2Vec module.  Four different formulas - 1) Cosine Similarity of Tweet Vector Sum (CSTVS) $1 - \frac{\alpha \cdot \sum_{i=1}^k \tau_i}{\lVert \alpha \rVert \lVert \sum_{i=1}^k \tau_i \rVert}$, 2) Dot Product of Search Term Vector and Tweet Vector Sum (DP) $\lVert \alpha \rVert \times \lVert\sum\limits_{i=1}^n \tau_i\rVert \times \cos{\theta}$, 3)  Mean Cosine Similarity (MCS) $\frac{1}{n}\sum\limits_{i=1}^n \cos(\theta_{\alpha}^{\tau_i})$, 4) Sum of Cosine Similarity over Square Root of Token Count (SCSSC) $\frac{1}{\sqrt{n}}\sum\limits_{i=1}^n \cos(\theta_{\alpha}^{\tau_i})$, were employed to score a tweet based upon its component word vectors. CSTVS is a programmatic implementation of the cosine distance formula \citep{salton_extended_1983} allows an efficient calculation of cosine distance. Cosine Similarity can be calculated by subtracting this value from 1. DP treats the sum of the vectors in a tweet as a vector itself ($\sum_{i=1}^k \tau_i$), and calculating the dot product of this interpreted vector and the vector for the search term ($\alpha$) returns a value that is proportional to the cosine similarity. MCS is the mean cosine similarity of the search term to all terms in a tweet, where $n$ is the number of terms in the tweet. SCSSC is similar in function to the MCS, where it reduces the impact of a shorter tweet by dividing by the square root of the count of tokens in a tweet ($n$). All formulas return a scalar score for a tweet - search term similarity match.

In order to evaluate the model, the codified data set of 19,000 tweets were used. The codification was done by a single human coder and a sampled set of tweets (100 with balanced distribution) was verified by two additional coders to access the inter-rater reliability. Tweets were labelled to be Irma \textit{related} if matched the following criterion --- 1) \textit{Explicitly contains references to ``Irma'' or ``Hurricane''}; 2) \textit{Contains current meteorological data, such as wind speed, rainfall levels, etc.}; 3) \textit{Refers to weather events such as storm, flood(ing) and rising water, rainfall, tornado, etc.}; 4) \textit{Describes the aftermath of extreme weather: trees down, power out, damage to buildings or construction, etc.}; 5) \textit{contains references to emotional states exacerbated by the weather: worrying about shelter, concerns for safety, pleas for help, etc.}; 6) \textit{Lists availability or absence of necessities: shelter, water, food, power, etc}. A message was labelled \textit{not related} if it met following criterion --- 1) \textit{mentioning a location absent any of the above content}; 2)\textit{Containing an attached picture that may be Irma related, but no additional text}; 3) \textit{Expressing emotions about the state of an event, but its connection to weather is ambiguous, i.e. a sporting event canceled, but no explanation as to why;} 4) \textit{Expressing emotions about a person’s condition, but its connection to weather is ambiguous: for ex: ``I hope \text{@abc123} gets better soon!''.} The resulting dataset had 8,296 tweets related to the Irma and 10,792 tweets not related. The inter-rater reliability of the codified messages using Light's Kappa metric was at $.69$, suggesting significant agreement between coders \cite{light1971measures, mchugh2012interrater}. 
 
This dataset was then used to evaluate the aforementioned formulas for different thresholds of the scores by analyzing the ratio of correctly classified tweets by the model. Hyper-parameters of the Word2Vec model were also tuned using the labeled tweets. The parameters selected for testing were context word window sizes from 1 to 10 words on either side of the target word; hidden layer dimensionality in 50D increments from 50D to 500D; minimum word occurrence from 0 to 9; negative sampling from 0 to 9 words. The cross product of the values contained in these ranges were used as the testing set of tuples for the training operations. For each set of parameters, the NN was trained through varying epochs, and the resultant word embeddings used in conjunction with the four scalar formulas to calculate scores for each tweet. The scores for each iteration were min-max scaled for the time delta, and the AU-ROC calculated based upon the thresholds of the scores in relation to the human-coded tweets.

\subsection{Overall model}
\label{sec:overall}

\begin{figure}[ht!]
\centering
\includegraphics[width=\textwidth]{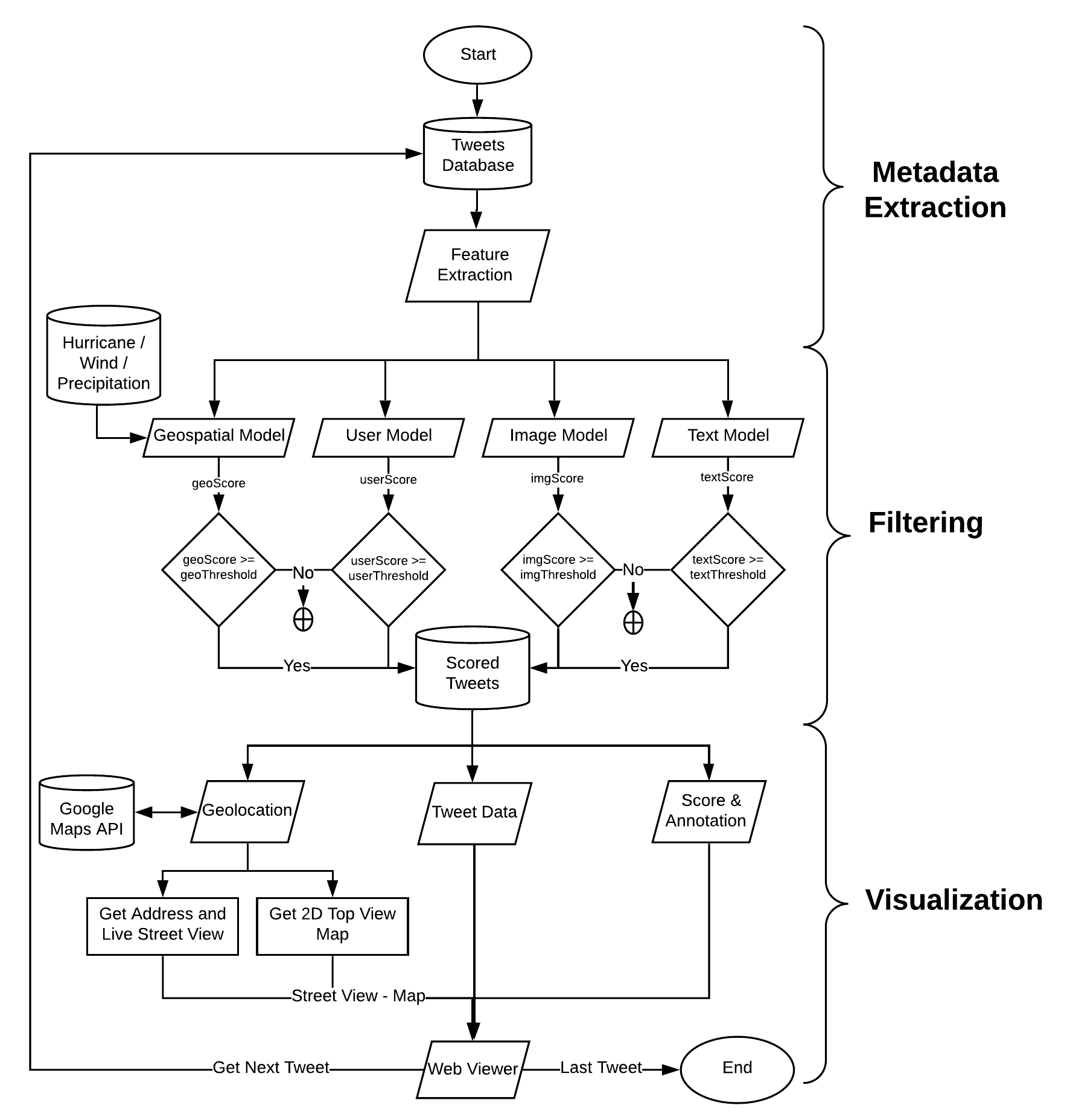}
\caption{Overall information flow model. The Metadata Extraction stage develops variables from the raw Twitter data, Filtering stage utilizes the developed 1) Geospatial, 2) User, 3) Image, and 4) Text analysis modules to score tweets, and the Visualization stage is used to observe the at location posted image along with Google Street View} 
\label{fig:overall_mod}
\end{figure}

Following the creation of individual models, we combined the results of each into a single overall model (Figure~\ref{fig:overall_mod}) which consists of three distinct stages -  1) Metadata extraction, 2) Filtering, and 3) Visualization of filtered tweets. For the first stage, the input is a tweet as a data-point. The metadata extraction stage mines the relevant attributes (image, geolocation, user, text) needed for the individual models of 1) Geospatial, 2) User, 3) Image, and 4) Text analysis. 

The results of the individual models are then combined in the second stage of filtering, where the normalized scores (decision score ranging from 0-100) for each models are combined at different thresholds to filter the relevant Twitter messages for Hurricane Irma. Any tweets without images are assigned an $imgScore = 0$, this allows users to view messages which contain images by setting the threshold to be $imgScore > 0$. The flexibility of the approach is in its ability to select different thresholds for respective models. This allows for a more generalizable model where a user can choose different set of thresholds for disparate disaster events. A logical $AND$ operation is used to obtain messages which pass all of the thresholds for each of the individual models. Specifically, a datapoint can only pass the filtering stage if all of its individual model scores are greater than or equal to the thresholds set. 

The filtered data are then stored in a database (Scored Tweets), where each datapoint can then be viewed on a visualization platform. The visualization platform extracts the location information from each datapoint (Geolocation), which is then cross-referenced with Google Maps API to provide three attributes --- 1) Google Street View \citep{anguelov2010google, rundle2011using}, 2) Physical address, and 3) A 2D top down view of the map at the location. These attributes (Street View - Map) along with the Tweet Data (text of the tweet, date-time, user, image, etc) and Score \& Annotation information ($\mathsf{P}(Related/Not-Related)$ and $\mathsf{P}(Tag)$, where $\mathsf{P}$ is the probability and $Tag \in \{Flooding, Windy, Destruction\}$) is then displayed on a web viewer. This presents an easy to use interface to view and visualize the messages for situational awareness. 

\section{Results}

\subsection{Geospatial}

\begin{figure}[ht!]
\centering
\includegraphics[width=\textwidth]{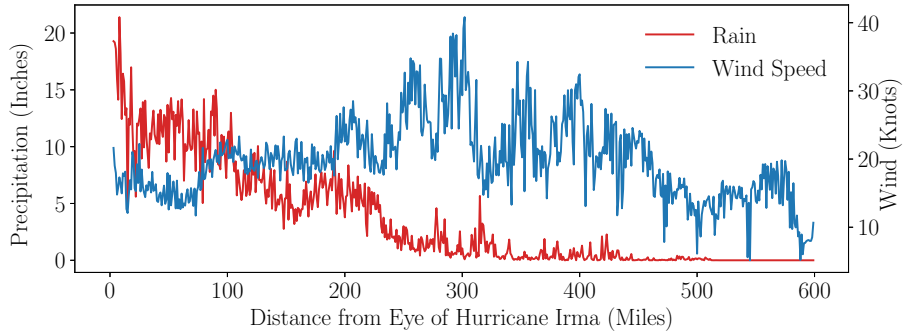}
\caption{Precipitation and wind speed in relation to the distance from Hurricane Irma’s eye.} 
\label{fig:rainwind}
\end{figure}

\begin{figure}[ht!]
     \centering
     \subfigure[Min-Max Normalization Scores.]{
        \resizebox*{5.75cm}{!}{\includegraphics{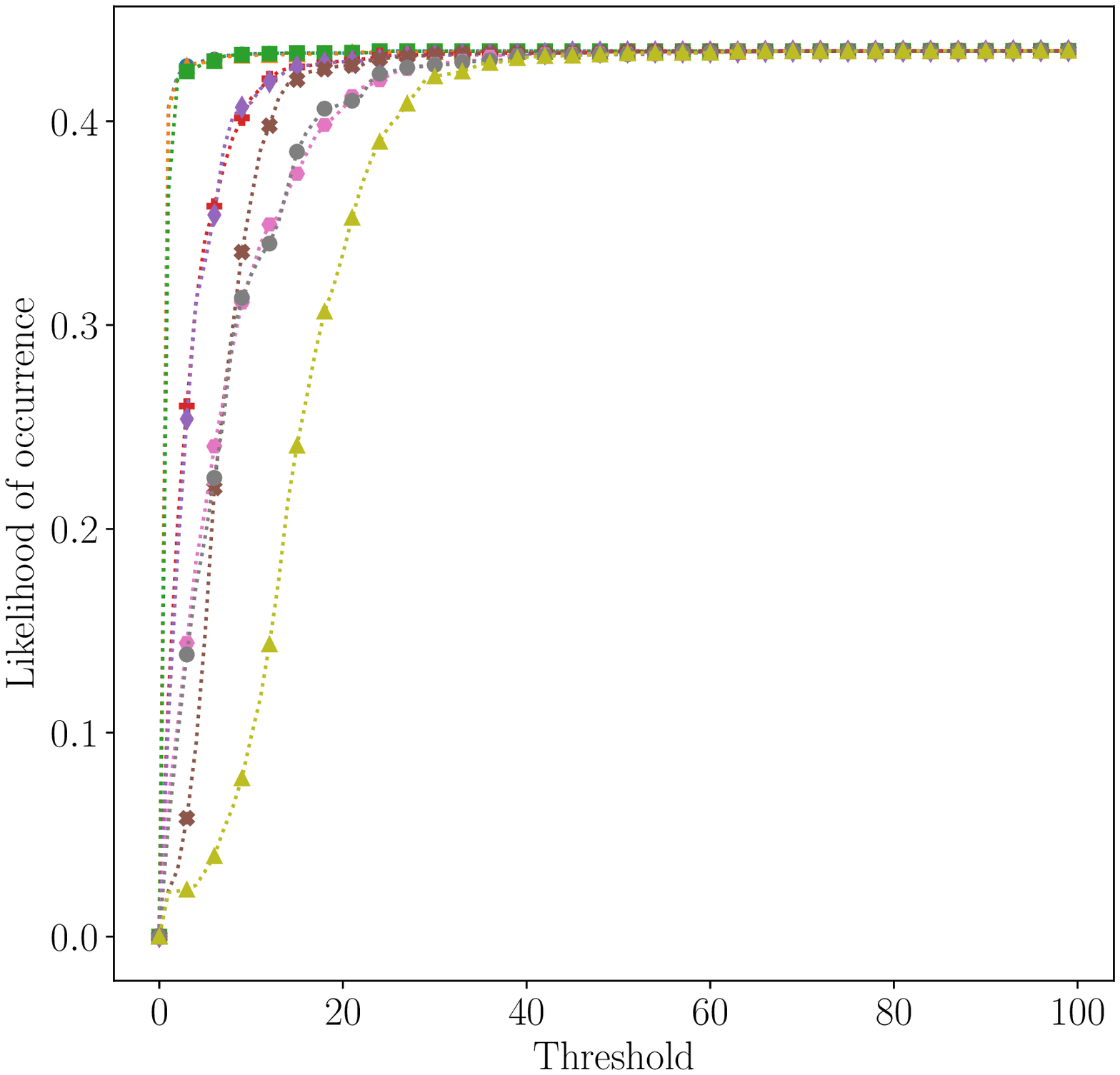}}}
    \subfigure[$log_{10}()$ Transformed Scores.]{
        \resizebox*{5.75cm}{!}{\includegraphics{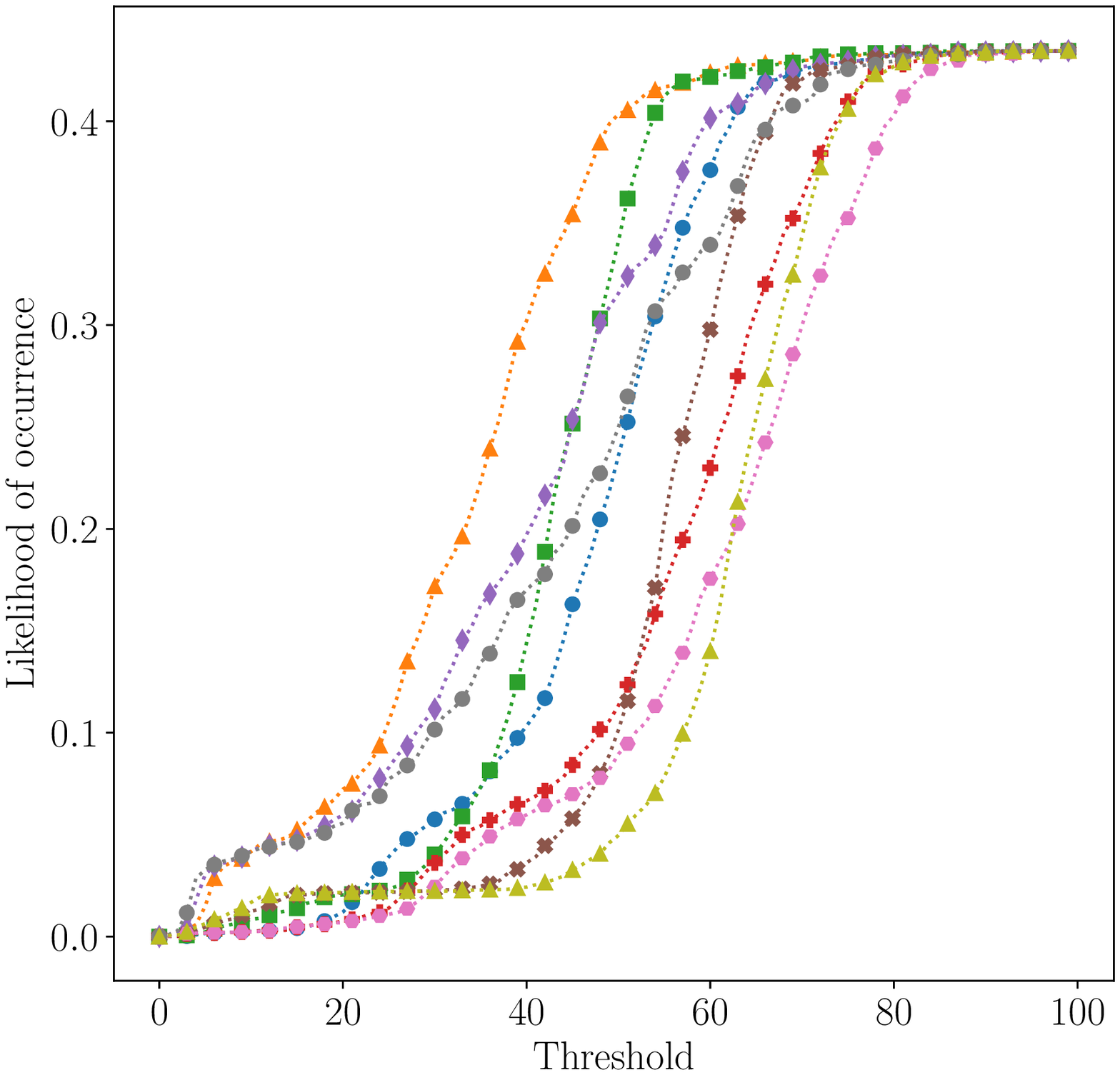}}}
    \subfigure[Box-Cox Transformed Scores.]{
        \resizebox*{5.75cm}{!}{\includegraphics{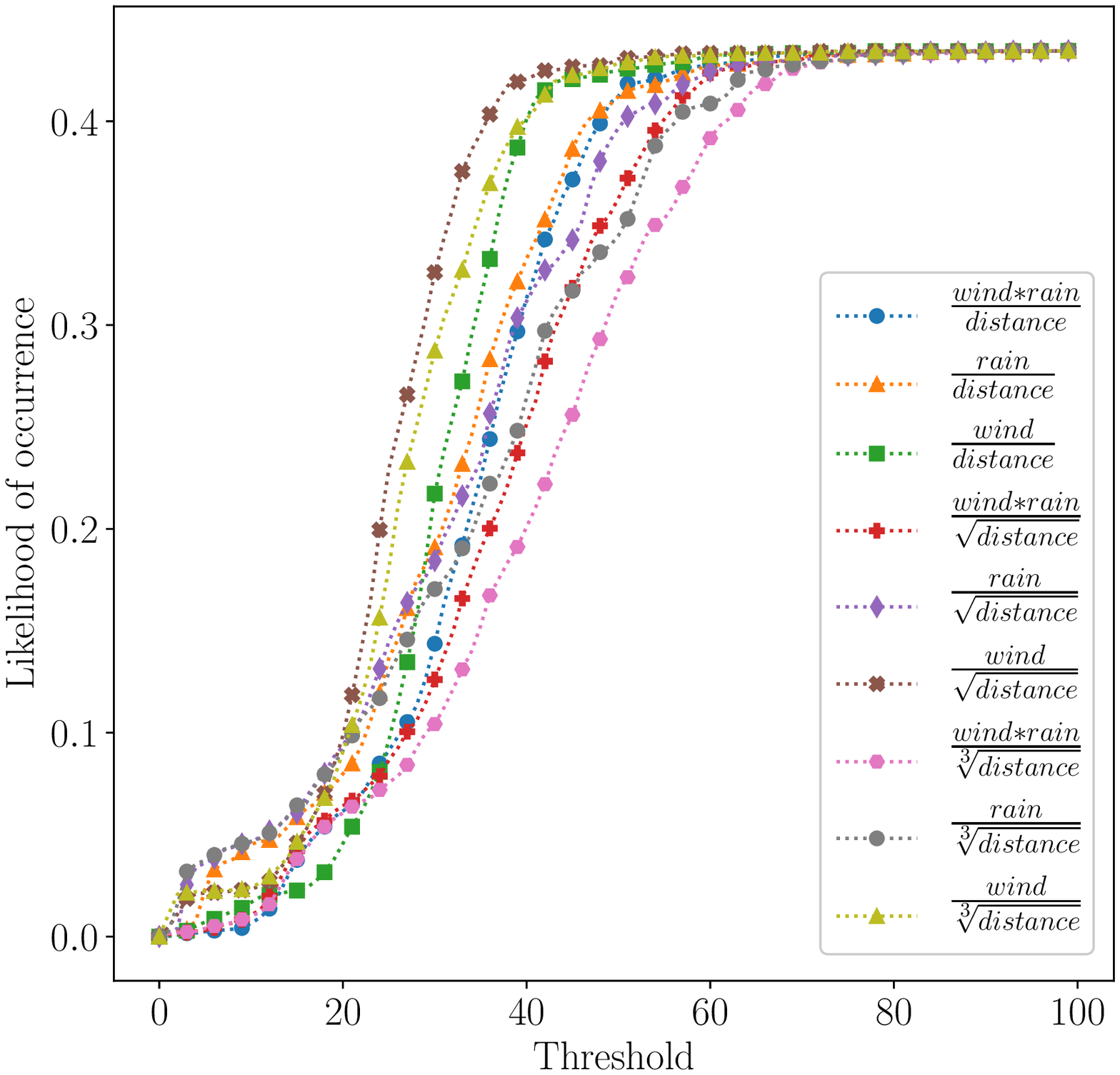}}}
    \caption{Cumulative Distribution Function (CDF) for Min-Max Normalization, Log, and Box-Cox transformed geospatial scores for the nine models. The common legend of all three figures is shown in Figure c.}
    \label{fig:cdf_plots}
\end{figure}

Preliminary exploration of the sensor readings for precipitation and wind speed with relative distance from the eye of the hurricane are shown in Figure~\ref{fig:cdf_plots}. Precipitation decreases exponentially farther away from the eye of the hurricane, measuring 5 - 20 inches. Median wind speeds have their peak around 300 miles from the eye of the hurricane.

\begin{table}[ht!]
\centering
\caption{Shapiro-Wilk statistics value for all the models . Top 5 values highlighted.}
{
\begin{tabular}{cc|l|l} \toprule
 & \multicolumn{3}{c}{\textbf{Normalization method}} \\ \cmidrule{2-4}
\textbf{Model}                   & $\frac{X-\min(X)}{\max(X)-\min(X)}$ & \textbf{$log_{10}()$} & $\gamma()$  \\ \midrule 
$\frac{wind\ *\ rain}{distance}$   & $0.07$   & $\boldsymbol{0.99}$ & $0.99$ \\[0.75ex] \hline
$\frac{rain}{distance}$   & $0.06$   & $0.92$ & $0.93$ \\[0.75ex] \hline
$\frac{wind}{distance}$   & $0.13$   & $0.95$ & $0.97$ \\[0.75ex] \hline
$\frac{wind\ *\ rain}{\sqrt[]{distance}}$ & $0.53$ & $\boldsymbol{0.98}$ & $\boldsymbol{0.98}$ \\[0.75ex] \hline
$\frac{rain}{\sqrt[]{distance}}$ & $0.51$ & $0.88$ & $0.89$ \\[0.75ex] \hline
$\frac{wind}{\sqrt[]{distance}}$ & $0.88$ & $0.88$ & $\boldsymbol{0.98}$ \\[0.75ex] \hline
$\frac{wind\ *\ rain}{\sqrt[3]{distance}}$ & $0.65$ & $\boldsymbol{0.98}$& $\boldsymbol{0.98}$ \\[0.75ex] \hline
$\frac{rain}{\sqrt[3]{distance}}$ & $0.65$ & $0.85$& $0.86$ \\[0.75ex] \hline
$\frac{wind}{\sqrt[3]{distance}}$ & $0.96$ & $0.85$& $\boldsymbol{0.99}$ \\[0.75ex] \bottomrule
\end{tabular}}
\label{tab:geomodel1}
\end{table}

Nine different geospatial models were developed and compared for their performance to filter Irma related tweets. Specifically, for each model the results calculated ratio of Irma related tweets, i.e. number of Irma related tweets / total number of tweets, at different thresholds between 0 and 1 (all values were min-maxed for normalization). Irma related tweets were identified by codification of 19,000 messages by human coder (annotation criteria described in Section \ref{sec:textmodel}). Figure 3, compares the cumulative distribution function (CDF) plots between the each of the functions within a subplot. The plots (a, b, c) further compare the results between - a. Min-Max Normalization, b. Log ($\log_{10}()$), and c.  Box-Cox ($\gamma()$)) transformation scores.

As observed, the results of the Log and Box-Cox transformations show a wider distribution of the ratio in comparison to the Min-Max normalized values, across the different thresholds. The results are also confirmed by the Shapiro-Wilks test (Table~\ref{tab:geomodel2}) where the Log and Box-Cox transformed models have higher scores, suggesting a more normal distribution of the results than the non-transformed ones. Based on the test the top five functions identified were - ($\gamma(\frac{wind}{\sqrt[3]{distance}})$, $\gamma(\frac{wind\ *\ rain}{distance})$, $log_{10}(\frac{wind\ *\ rain}{distance})$, $\gamma(\frac{wind\ *\ rain}{\sqrt[]{distance}})$, and $log_{10}(\frac{wind\ *\ rain}{\sqrt[]{distance}})$. Each of the models were in very close proximity to the scores observed in the test.   

\begin{figure}[ht!]
\centering
    \subfigure[CDF Scores.]{
         \resizebox*{8cm}{!}{\includegraphics{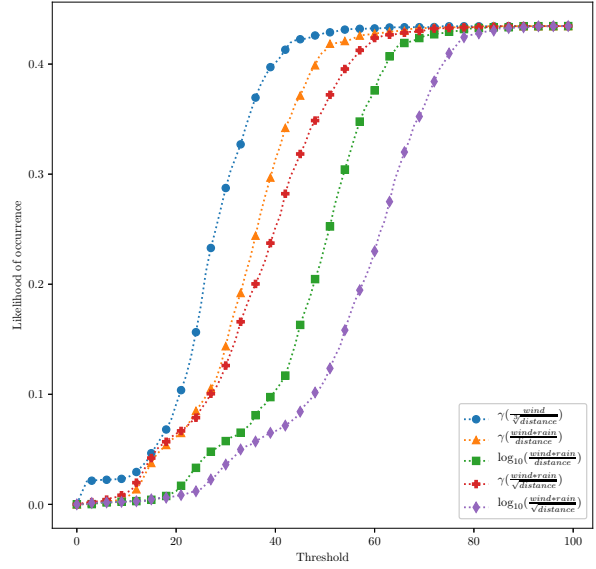}}}
     \subfigure[F1 Scores.]{
        \resizebox*{8cm}{!}{\includegraphics{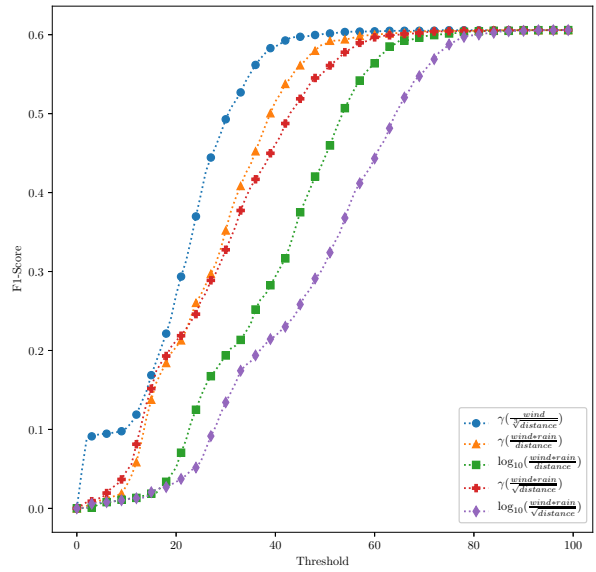}}}
\caption{Cumulative Distribution Function (CDF) and F1-Scores for the top five \textit{geospatial} models.} 
\label{fig:cdfgeo}
\end{figure}

Additional analysis was conducted to observe the statistical properties of the top five models.  Figure~\ref{fig:cdfgeo} shows the CDF and F1-Scores for each of these functions. Table~\ref{tab:geomodel2}, show the general statistical properties. Out of the five, $log_{10}(\frac{wind\ *\ rain}{\sqrt[]{distance}})$ was chosen as a final model function, based on its mean being the closest to 0.5. 

\begin{table}[ht!]
\centering
\caption{General data statistics for top 5 models. $log_{10}(\frac{wind\ *\ rain}{\sqrt{distance}})$ was selected as the normalization model for geospatial analysis as its distribution mean was closest to $0.5$}
{
\begin{tabular}{cc|c|c|c} \toprule
 & \multicolumn{4}{c}{\textbf{Data Statistics}} \\ \cmidrule{2-5}
\textbf{Model}                   & Shapiro-Wilks & \multirow{2}{*}{\makecell{Standard \\ Deviation ($\sigma$)}} & \multirow{2}{*}{\makecell{Mean \\ $\mu$}} & \multirow{2}{*}{\makecell{\% of Data within \\ $-1 \leq \sigma \leq 1$}} \\[3ex] \midrule 
$\gamma(\frac{wind}{\sqrt[3]{distance}})$   & $0.99$   & $0.12$ & $0.28$ & $0.66$ \\[0.75ex] \hline
$\gamma(\frac{wind\ *\ rain}{distance})$   & $0.99$   & $0.14$ & $0.38$ & $0.65$ \\[0.75ex] \hline
$log_{10}(\frac{wind\ *\ rain}{distance})$   & $0.99$   & $0.14$ & $0.39$ & $0.65$ \\[0.75ex] \hline
$\gamma(\frac{wind\ *\ rain}{\sqrt{distance}})$   & $0.98$   & $0.16$ & $0.43$ & $0.64$ \\[0.75ex] \hline
$\boldsymbol{log_{10}(\frac{wind\ *\ rain}{\sqrt{distance}})}$   & $\boldsymbol{0.98}$   & $\boldsymbol{0.16}$ & $\boldsymbol{0.46}$ & $\boldsymbol{0.64}$ \\[0.75ex] \hline
\end{tabular}}
\label{tab:geomodel2}
\end{table}

\subsection{Image classification}

\begin{figure}[ht!]
\centering
\includegraphics[width=.6\textwidth]{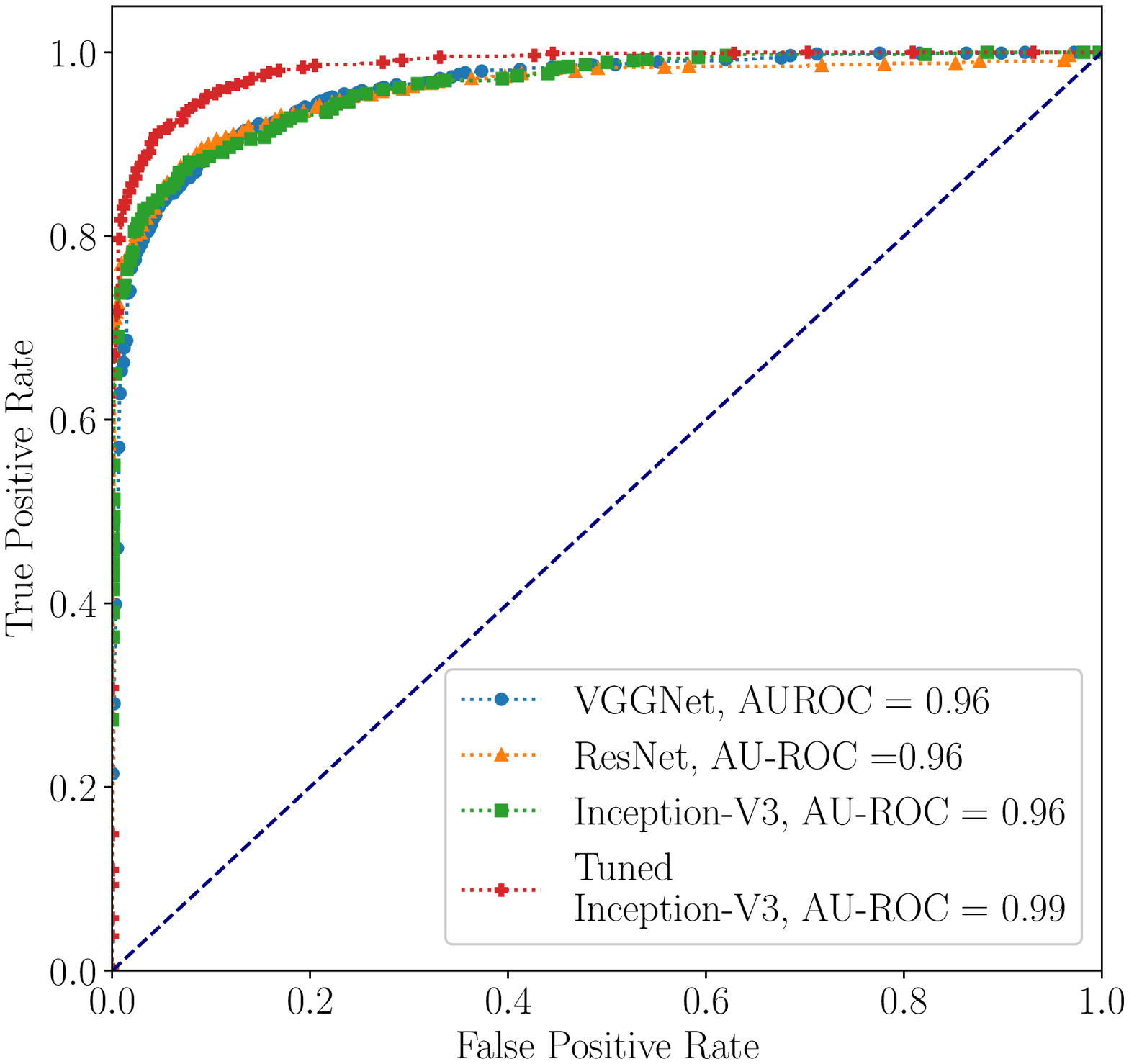}
\caption{Area Under - Receiver Operating Characteristics (AU-ROC) Curves for V3, VGG net, ResNet architecture and Tuned Inception V3 models for binary classification of \textit{images} (hurricane related versus non-hurricane related).} 
\label{fig:binaryauroc}
\end{figure}

The performance of various image classifiers are shown in Table~\ref{tab:imgmodel}. In the first stage of classification, which uses a binary classifier distinguish hurricane and non-hurricane related images, the Tuned Inception V3 architecture performed the best with an overall F1-score of 0.962. Figure~\ref{fig:binaryauroc}, shows the comparative AU-ROC curves for the different models. Between the classes, the Tuned Inception V3 model also performed well with an F1-score of 0.959 for class 1 (hurricane related) and 0.965 for class 0 (non-hurricane related) images. 

\begin{table}[ht!]
\centering
\caption{Performance comparison of deep-learning models (Inception-V3, VGGNet, ResNet, and Tuned Inception-V3) for binary classification and multi-label annotation.}
{
\begin{tabular}{cc|c|c|c|c|c} \toprule
\multirow{4}{*}{\textbf{Model}} & \multicolumn{6}{c}{\textbf{Performance Measures}} \\ \cmidrule{2-7}
                                & \multicolumn{3}{|c|}{\textbf{Binary Classifier}}  & \multicolumn{3}{c|}{\textbf{Multi-Label Annotator}} \\ \cmidrule{2-7}
                                
    & \multicolumn{1}{|c}{Precision} & Recall & F1-Score & Precision & Recall & \multicolumn{1}{c|}{F1-Score} \\[0.75ex] \midrule 
VGGNet   & $0.88$   & $0.87$ & $0.88$ & $0.70$ & $0.60$  & $0.64$\\[0.75ex] \hline
ResNet   & $0.88$   & $0.89$ & $0.89$ & $0.68$ & $0.61$  & $0.64$\\[0.75ex] \hline
Inception-V3   & $0.89$   & $0.88$ & $0.88$ & $0.75$ & $0.72$ & $0.73$\\[0.75ex] \hline
\multirow{2}{*}{\makecell{\textbf{Tuned} \\\textbf{Inception-V3}}}   & \multirow{2}{*}{$\boldsymbol{0.96}$}   & \multirow{2}{*}{$\boldsymbol{0.95}$} & \multirow{2}{*}{$\boldsymbol{0.95}$} & \multirow{2}{*}{$\boldsymbol{0.90}$} & \multirow{2}{*}{$\boldsymbol{0.92}$} & \multirow{2}{*}{$\boldsymbol{0.91}$}\\[3.2ex] \hline
\end{tabular}}
\label{tab:imgmodel}
\end{table}

\begin{figure}[ht!]
\centering
\includegraphics[width=.6\textwidth]{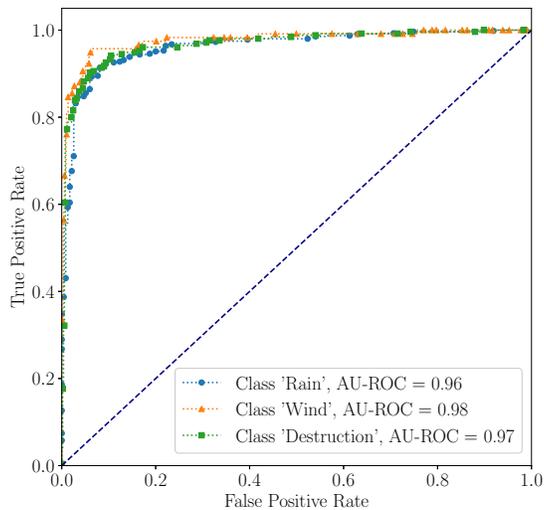}
\caption{Area Under - Receiver Operating Characteristics (AU-ROC) Curves for Tuned Inception V3 model for multi-label annotation for \textit{images} - 1) 'Flood', 2) 'Wind', and 3) 'Destruction'.} 
\label{fig:multiauroc}
\end{figure}

The hurricane related images were then fed through a second round of classification trained on multi-label annotation of - 1) flood, 2) wind, and 3) destruction. Table~\ref{tab:imgmodel} also compares the results of the analysis, where the Tuned Inception V3 architecture outperformed the other models, with an average F1-score of 0.896. Within the classes, the F1-scores were well distributed with class 1) as 0.821, 2) as 0.888, and 3) 0.941. Figure~\ref{fig:multiauroc} shows the AU-ROC curves for the different annotations performed on the images by the Tuned Inception V3 architecture.

Analyzing the cutoff thresholds of the probability scores for the Tuned Inception V3 model, shows a distribution with a mean of 0.63, a median of 0.75, and a standard deviation of 35.07. The values show a wide distribution of probability scores, which is useful in having a wider range in the cutoff thresholds used for filtering the images. 

\subsection{User}

\begin{figure}[ht!]
\centering
\includegraphics[width=.6\textwidth]{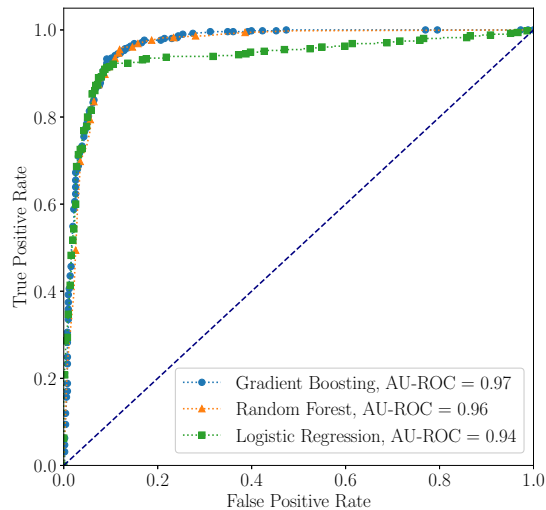}
\caption{AU-ROC Curves for Random Forest, Gradient Boosted, and Logistic Regression Classifiers in predicting Verified \textit{users}.} 
\label{fig:userauroc}
\end{figure}

The F1-score of the Random Forest (RF), Gradient Boosted (GB), and Logistic Regression (LR) models  of the models trained on predicting user verification were recorded at 0.97, 0.92, and 0.88 respectively. Figure~\ref{fig:userauroc} shows the comparative AU-ROC scores of the different models, where the RF classifier is able to outperform the rest of the models. The best performing RF model was developed by using a grid search approach, where multiple model parameters (number of estimators, depth, leaf splits, etc.) were evaluated.  The resulting model had a precision, recall, and AU-ROC were observed to be 0.96, 0.98, and 0.99 respectively.

The classifier was balanced in its prediction accuracy in both verified (class 1) versus non-verified (class 0) users (Figure~\ref{fig:user_randomf}). The output probability values of the binary model were further min-maxed to a threshold score between 0 and 100. The resulting normal distribution had a mean of 50.56, a median of 66.26, and a standard deviation of 39.69.

\subsection{Text}
\begin{figure}[ht!]
\centering
\includegraphics[width=.6\textwidth]{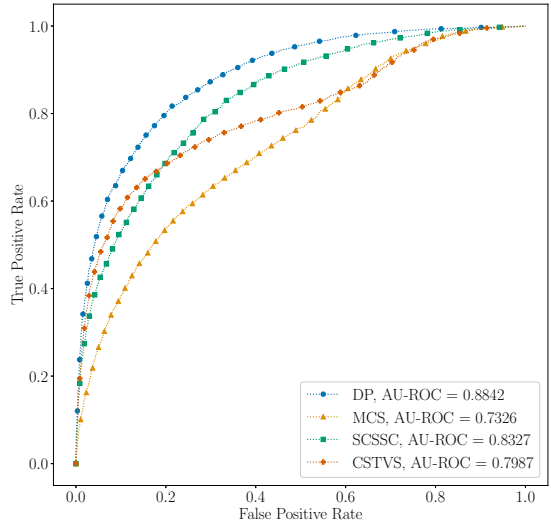}
\caption{AU-ROC Curves for \textit{text} --- 1) Cosine Similarity of Tweet Vector Sum (CSTVS) , 2) Dot Product of Search Term Vector and Tweet Vector Sum (DP) , 3)  Mean Cosine Similarity (MCS), 4) Sum of Cosine Similarity over Square Root of Token Count (SCSSC).} 
\label{fig:text_auc}
\end{figure}

\begin{figure}[ht!]
     \centering
     \subfigure[Confusion matrix of non-verified (class 0) versus verified (class 1) user prediction with Random Forest Classifier.]{
        \resizebox*{6.5cm}{!}{\includegraphics{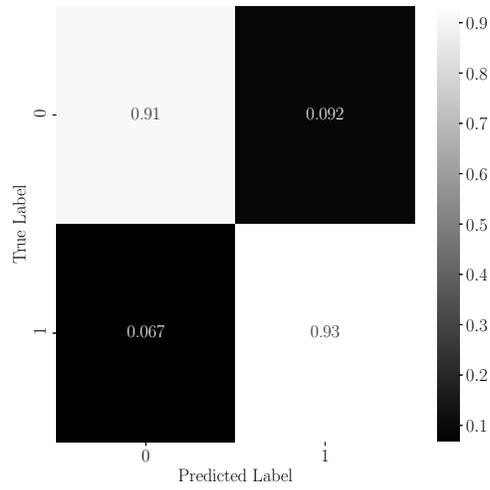}}}\hspace{5pt}
    \subfigure[Relative importance between features (using information gain) for prediction verified users using Random Forest Classifier.]{
        \resizebox*{6.5cm}{!}{\includegraphics{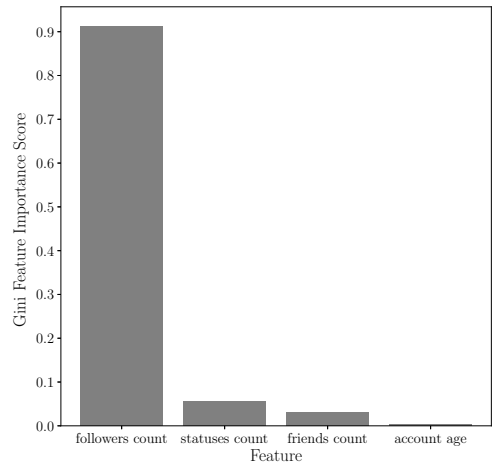}}}
    \caption{Performance of the \textit{user} Random Forest Classifier in the binary classes and feature importance metrics. }
    \label{fig:user_randomf}
\end{figure}

The results of the text analysis module were based on the binary categorization of the tweets codified as ‘irma related’ (class 0) or ‘non irma related’ (class 1). Evaluation of the four different resulted in the F1-scores of .6553 - MCS, .7824 - DP, .7049 - CSTVS, and .7347 - SCSSC. We observe the dot product between search term vector and tweet vector sum (DP) gives us the best result. Figure~\ref{fig:text_auc} shows the AU-ROC curves comparing the different formula performance in the analysis.

Each model was further evaluated to identify the best set of parameters. Within the analysis we found the DP formula was still the best performing with a word window size of 1, hidden layer dimensionality of 150, a minimum word count of 5, a negative sampling value of 1, and training the Word2Vec model through 25 epochs. The resulting normal distribution had a mean of 24.73, a median of 21.64, and a standard deviation of 14.05.

\section{Discussion}

We address each individual model separately before discussing the final combined model and providing limitations/ future directions for this work. 

\subsection{Individual models}

\subsubsection{Geospatial}

The geospatial models developed in the study provide a measure of relevance to a tweet by including the forcing sensor data (wind speed, precipitation, and distance from the eye of the hurricane). The best performing function of $log_{10}(\frac{wind\ *\ rain}{\sqrt[]{distance}})$ combines the values into a single normalized score which can be used to weight a geographic/sensor relevancy factor for any tweet.  More specifically, the function helps us identify Twitter messages at locations which are in close proximity to the hurricane forcing and have observed increased amount of precipitation and wind speed. As seen in the results, the chosen $log_{10}(\frac{wind\ *\ rain}{\sqrt[]{distance}})$ was the closest to a normally distributed function. This allows for a greater granularity on threshold cutoff points in comparison to other functions, leading to a fine-grained control over filtering based on the geographic relevance of the tweets. The statistical properties of the function also enables analysis of confidence intervals which can be used to ascertain the reliability of a message within the context of sensor data. In other words, tweets with anomalous sensor readings can be easily identified, leading to more reliable mining of messages related to the disaster event.

We envision that filtering tweets using their geospatial information relative to storm position and also environmental factors can help isolate tweets from heavily impacted locations. By examining locations close to the storm, with high wind gusts, or heavy precipitation allows users to quickly examine locations that might be expected to show the most severe imapcts from storm events.  

\subsubsection{Image}

Comparing the performance of the CNN architectures (Inception V3, VGG, ResNet, and Tuned Inception V3) for binary classification (hurricane and non-hurricane related), we observe that the Tuned Inception V3 model (F1-score 0.95) has almost a 6-7\% accuracy gain over others. In comparison to the VGG and ResNet architectures, the Tuned Inception V3 larger number of parameters which can be trained to observe the nuances between the images. While the base Inception V3 classifier contains the same number of parameters, re-tuning the weights to our training sample of images improved its accuracy considerably for the binary classification. This can be attributed to the pre-training and transfer learning of the model, where it already had prior weights based on classification of physical objects, and our image data tuned it further for disambiguating physical and non-physical scenes.

We do observe a slight performance decrease (F1-score 0.91) of the architecture trained on the multi-label annotation of the images. This can be attributed to the limited number of training samples that were available to the classifier. The complexity of the images in the samples further degrades the performance, for example, images of lakes and sea water are not much different from images of flooding.

Prior research in the area of automating image analysis (using machine learning) from social media has primarily focused on quantifying the level of damage in disaster situations \citep{lagerstrom_image_2016,nguyen_damage_2017,li_localizing_2018}. Our approach uses a dual stage model, where the first stage is responsible for increasing the quality of images by filtering out the non-relevant / non-physical images. The output is then fed into the second stage for categorization into different groups based on situational conditions (flooding, wind, and destruction). While prior studies have looked at disambiguating “fake/altered” images \citep{gupta_faking_2013,marra_detection_2018}, they are based on analyzing the content of the tweet along with user reliability measures for training machine learning models. Within our approach we only utilize the image features for the training our models. The output is image scores are based on normalized probability values, which can be used for threshold cutoffs, where setting a high threshold will only mine the most hurricane related images. The second stage then annotates the images for further filtering of images based on the needs of the domain.Filtering images permits users to quickly focus on a small subset of visual information that is presumed to be most valuable for storm impact assessment, compared to needing to scroll through many images to find useful information. 

\subsubsection{User}
Prior studies \citep{buntain_automatically_2017,zhou_fake_2018,masood_spammer_2019,del_vicario_polarization_2018} focused on identifying incorrect/fake/altered information in social media have established the source of information (social media user) as a key component. A large proportion of the studies \citep{hutchison_detecting_2010,subrahmanian_darpa_2016,efthimion_supervised_2018,sahoo_hybrid_2019} have been based on developing machine learning approaches towards detection of “bots” or fake user accounts \citep{ferrara_rise_2016} on social media. For example, \cite{karami_twitter_2019} identify the credibility of the user as an important element in mining good quality situational awareness information from social media. Within our approach, we leverage prior work done in the field by identifying the user features of account age, status count, number of followers, number of friends, existence of url links, number of hashtags, existence of images, retweets, geolocation, and  message frequency in training our machine learning models.

Comparing the results between the parametric (Logistic Regression) and the non-parametric ensemble models (Random Forest and Gradient Boosting), we observe the ensemble models are able to outperform by a margin of 4-7\%. The developed Random Forest model has a very high accuracy (F1-score 0.97, AU-ROC 0.99) in disambiguating between verified and non-verified users. While the ratio of the number of verified versus non-verified users was imbalanced (approximately 1:100) in our data, the developed RF model is able to accurately distinguish between the classes as shown by the confusion matrix (Figure 8). 

Further analyzing the RF model, we calculated the average decrease in Gini impurity/information gain (entropy) among all estimators to observe the importance of features. Specifically, as estimators are developed on a subset of features, the decrease in information gain across a subset of features can be used to infer the relative importance of features. Figure 9, shows the relative importance of four features (rest where too low to observe), where the number of followers, status, and friends, along with the account age are the top features which affect the decision of the model towards the credibility of a Twitter user in our data.

The analysis of features within our model shows similar feature importance measures to that used in prior research to identify reliable information sources \cite{karami_twitter_2019}. However, our approach provides a more generalized model where a thresholding on probability scores can be used to select user sources based on needs of a specific event. The approach is also dynamic where a model can be quickly retrained using the available ''Verified'' tags instead of manually re-annotating accounts. This prevents temporal dilation of features where a model trained on an older labeled dataset cannot perform as well due to the changes in account statistics over time. 


\subsubsection{Text}

With the observed dot-product based model performing the best with the F1-score analysis, we applied the model towards an hourly aggregated corpus within our data. Specifically, when the corpus was confined to the tweets from a single hour, the vector representations of word embeddings were only influenced by the contexts derived from that hour. Words would have a unique vectorization specific to that hour, and relationships between words were dependent on the context interpretations within that time. The cosine similarity of two terms could be calculated for this duration, and words with the highest scoring cosine similarity to a term would indicate an observed relationship that was finite within the timeframe. In short, two words could be similar in one hour, and completely different the next, depending on the content of the tweets at the time.

Table~\ref{tab:words} shows the output of the DP model for the hourly aggregated tweets. Prior to landfall (time 13:00), we observe mentions of the ``storm'', ``wind'', ``eye'', ``ese'' (East-South-East), ``e'' (East), etc, having prominence in the top 20 words as identified by the DP model to be semantically similar to search term ``irma''. There is consistency in the thematic representation where these words did occur across the 6 hours prior to the hurricane. During the window of the hurricane Irma’s landfall we observe ``shelter'', ``\#hurricaneirma'', ``eye'', ``landfall'', ``help'', ``plea'', etc., as the most related terms to ``irma''. After the hurricane the context of the ``irma'' changes to reflect more help/rescue/concern words where ``shelter'', ``safe'', ``check'', ``food'', ``power'', etc. become the most prominent words.

\definecolor{dk-before}{HTML}{ffb399}
\definecolor{md-before}{HTML}{C04C5C}
\definecolor{lt-before}{HTML}{D3828D}

\definecolor{dk-during}{HTML}{95711E}
\definecolor{md-during}{HTML}{BB9337}
\definecolor{lt-during}{HTML}{CFB373}

\definecolor{dk-after}{HTML}{ace600}
\definecolor{md-after}{HTML}{5A7DB4}
\definecolor{lt-after}{HTML}{8BA4CA}

\begin{table}[ht!]
\centering
\caption{Hourly aggregate of top 20 semantically related to terms to “Irma”, for six hours prior and after landfall. The words have been stemmed to their root. \#hirma denotes the hashtag \#hurricaneirma used in the tweet. Colors indicate similar terms across the different time windows of the hurricane.}
{
\resizebox{\textwidth}{!}{
\begin{tabular}{clllllllllllll}\toprule
\multirow{3}{*}{\makecell{\textbf{Word} \\ \textbf{Rank}}} & \multicolumn{13}{|c|}{\textbf{Time}} \\ \cmidrule{2-14}
    & \multicolumn{1}{|l}{\textbf{7:00}}    & \textbf{8:00}      & \textbf{9:00}         & \textbf{10:00}      & \textbf{11:00}     & \textbf{12:00}        & \multirow{2}{*}{\makecell{\textbf{13:00} \\ \textbf{Irma Landfall} }}           & \textbf{14:00}   & \textbf{15:00}           & \textbf{16:00}    & \textbf{17:00}           & \textbf{18:00}           & \multicolumn{1}{l|}{\textbf{19:00}} \\ [3ex] \cmidrule{1-14} 

1.  & sleep   & last      & \cellcolor{md-before}ese          & \cellcolor{md-before}ese        & tampa     & tampa        & \cellcolor{dk-during}\textbf{shelter}         & \cellcolor{dk-during}shelter & \cellcolor{dk-during}shelter         & \cellcolor{dk-during}shelter  & \#hirma & \cellcolor{lt-before}hit             & tampa               \\
2.  & offici  & outsid    & outsid       & tri        & yet       & time         & \textbf{first}           & whole   & tampa           & want     & outsid          & \cellcolor{dk-after}safe            & \cellcolor{lt-during}eye                 \\
3.  & need    & sleep     & moder        & help       & time      & \cellcolor{lt-after}check        & \textbf{wait}            & tampa   & beauti          & time     & food            & outsid          & bay                 \\
4.  & \cellcolor{md-before}ese       & \cellcolor{md-before}ese          & beauti          & yet     & see             & good     & \textbf{get}             & open            & first               \\
5.  & want    & \cellcolor{md-before}e         & sleep        & outsid     & \cellcolor{lt-during}eye       & tri          & \textbf{tri}             & \cellcolor{lt-after}check   & \#hirma & tampa    & \cellcolor{dk-after}safe            & updat           & time                \\
6.  & \cellcolor{lt-before}hit     & \#key     & nation       & \cellcolor{lt-during}eye        & friend    & might        & \textbf{make}            & open    & prep            & get      & watch           & hurrican        & wait                \\
7.  & \#key   & wind      & \cellcolor{md-before}e            & moder      & first     & night        & \textbf{could}           & \cellcolor{lt-before}hit     & food            & guess    & time            & make            & \cellcolor{lt-before}hit                 \\
8.  & wind    & tropic    & need         & sleep      & night     & first        & \textbf{made}            & get     & come            & hurrican & peopl           & prayer          & outsid              \\
9.  & tropic  & good      & wind         & heavi      & last      & close        & \textbf{see}             & friend  & watch           & last     & know            & first           & \#hirma2017 \\
10. & see     & beach     & fuck         & wellington & close     & coffe        & \cellcolor{lt-during}\textbf{eye}             & read    & yet             & \cellcolor{lt-after}check    & see             & get             & make                \\
11. & much    & \#irma    & \#sfltraffic & wind       & \#traffic & friend       & \textbf{\#hirma} & world   & time            & \cellcolor{lt-before}hit      & love            & wait            & food                \\
12. & \#irma  & florida   & pleas        & fuck       & strong    & help         & \textbf{world}           & \cellcolor{dk-after}safe    & sleep           & peopl    & power           & everyon         & us                  \\
13. & beach   & \cellcolor{dk-before}storm     & \cellcolor{dk-before}storm        & tropic     & outsid    & last         & \textbf{night}           & good    & ride            & come     & gonna           & \cellcolor{lt-after}check           & \cellcolor{dk-during}shelter             \\
14. & florida & \#mfl     & beach        & good       & make      & follow       & \textbf{peopl}           & time    & first           & \cellcolor{lt-during}eye      & \#irma          & see             & get                 \\
15. & \cellcolor{dk-before}storm   & aso       & peopl        & see        & well      & outsid       & \textbf{close}           & come    & get             & friend   & still           & home            & last                \\
16. & know    & lauderdal & \#irma       & pleas      & want      & make         & \textbf{outsid}          & make    & \cellcolor{lt-after}check           & food     & hurrican        & power           & point               \\
17. & \#mfl   & power     & florida      & \cellcolor{dk-before}storm      & phone     & sleep        & \textbf{help}            & first   & go              & day      & \#nfl           & \#hirma & \cellcolor{dk-after}safe                \\
18. & power   & mesonet   & f            & flood      & sleep     & strong       & \cellcolor{dk-before}\textbf{landfal}         & beauti  & know            & see      & make            & okay            & open                \\
19. & call    & rain      & rain         & beach      & \cellcolor{lt-before}hit       & \#irmageddon & \textbf{come}            & wait    & open            & make     & home            & watch           & alway               \\
20. & aso     & \cellcolor{dk-after}safe      & mesonet      & rain       & florida   & open         & \textbf{pleas}           & home    & tri             & way      & want            & yet             & video \\             \hline
\end{tabular}}}
\label{tab:words}
\end{table}

The results show that the word embedding based dot product model is capable of identifying tweets which are most relevant to the search/seed term. This is highlighted by the example of the term ``ese'', which when taken by itself, might reference an informal Spanish colloquialism for ``man''. When interpreted within the hourly-divided corpora within this dataset, it takes on a different semantic interpretation. For the tweets occurring within each of the four hours immediately preceding landfall, ``ese'' is in the top twenty most related terms to ``irma'', and does not appear in the hourly lists following. Looking at the terms related to ``ese'' it can be determined that this refers to the abbreviation for East-South-East, likely referencing the direction from which the hurricane approached. After landfall, this term was no longer as relevant, and therefore less likely to appear as a related term. 

\subsection{Overall model}

\begin{figure}[ht!]
\centering
\includegraphics[width=.9\textwidth]{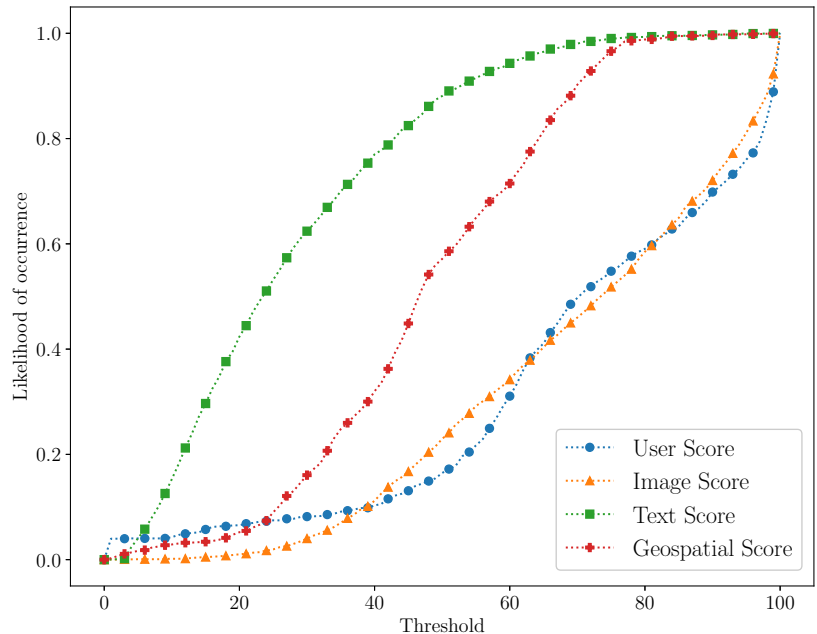}
\caption{CDF of Overall Model and percentage of tweets passing different model thresholds.} 
\label{fig:overall_cdf}
\end{figure}

In the overall model, the number of possible combinations for the thresholds is large (at $100^4$), where each of the four models can have a value ranging from 0-100. A cumulative distribution plot (CDF) was used to analyze the percentage of data-points passing the thresholds set for each of the models. Figure~\ref{fig:overall_cdf} shows the comparative analysis of each of the model,  where the curves are inversely proportional to the thresholds indicating a decrease in the percentage of tweets passing higher thresholds. Within the analysis, low thresholds are representative of more reliable sources and related contents, resulting in a low percentage of overall tweets passing through the filter. Similarly, at the threshold of 100, all tweets pass the filter providing complete access to all data.

The CDF plot also highlights the comparative performance of various models, where the text based filter includes a higher percentage of tweets at lower thresholds while the user verification filter includes most users at higher thresholds. Image classification also results in a similar performance to user verification (including most users at higher thresholds), whereas filtering based on geospatial scores filters more linearly. We observe high quality results (low false positives) at a likelihood occurrence of $0.6$. by setting initial thresholds to $30$ for text, $50$ for geospatial, and $85$ for both image and user scores. These recommended thresholds for Hurricane Irma provide a baseline for comparison with different events , and for hurricanes in different locations. 



\subsection{Limitations and Future Work}
Our current work explores the utilization of multiple modalities present in social media data to filter hazard event related information. We acknowledge certain limitations of this approach. Our approach is to cumulatively evaluate the operation of all sub-models in capturing the messages. As a result, we focused in this work on tweets that have all attributes: geolocation, text, and image (note all tweets have user attributes). However a smaller overall model with specific combinations of the sub-models can be used in certain conditions. For example, researchers who are interested in just messages with text can use an overall model that excludes the image sub-model and subsequently not filter based on a threshold for images.

Furthermore, our models are evaluated using the data from a single event — (Hurricane Irma) and a single location (Florida, USA). As a part of our future effort we plan to extend this framework to other hurricane events (and locations), such as, Maria, Harvey, and Florence, along with application of the approach to other disaster scenarios, such as fires, earthquakes, floods, etc. to aid in understanding the filtering step and thresholds in other contexts. Each event will likely have different specifications on the quality of data that needs to be extracted, for which we need to cross evaluate the approach against various events to provide recommendations for thresholds to be used for different disaster categories.

Our approach can operate as a primary filtering mechanism for additional anlysis to extracting information during a disaster event. Additional models which help with categorization of messages, such as, disaster damage quantification, information, requests of help, resources offerings, organizing efforts, etc., can be implemented to extract higher level information from the data. 

\section{Conclusion}
In this study, a multimodal filtering approach was developed and evaluated to extract and subset geocoded images posted on Twitter within the context of Hurricane Irma. Our prototype model consisted of four sub-models: geospatial, image classification, user credibility, and text analysis. Each sub-model returned a score in the range of 0-100 and allowed for user-defined filtering based on bespoke thresholds. Each of the four models aim to filter information about reliability, information consistency, and overall usefulness of the message. This single combined model shows potential for application in disaster and emergency contexts, allowing users to quickly search and filter for relevant geolocated tweets.

\section{Acknowledgments}

The authors would like to thank UNCG undergraduate students Kaitlyn Jessee and Elaina Kauzlarich for classifying the images associated with the tweets. This study was supported in part by the National Science Foundation (Awards\# CMMI-1541136 and \# SES-1823633), the Eunice Kennedy Shriver National Institute of Child Health and Human Development (Award \# P2C HD041025), DoD/DARPA (R0011836623/HR001118200064), seed grant initiatives from UNC Greensboro and Penn State's Social Science Research Institute, Population Research Institute, and Institute for CyberScience, and an Early-Career Research Fellowship from the Gulf Research Program of the National Academies of Sciences, Engineering, and Medicine. The content is solely the responsibility of the authors and does not necessarily represent the official views of the Gulf Research Program of the National Academies of Sciences, Engineering, and Medicine.

\section{Data and codes availability statement}

The data and the codes used in the research are available on Figshare: \url{https://figshare.com/s/235146fc2d6de33654f3} \citep{datacite}



\section{References}
\bibliography{bare_jrnl.bib}

\end{document}